\documentclass[%
 reprint,
 amsmath,amssymb,
 aps,
prb,
]{revtex4-2}

\usepackage[whole]{bxcjkjatype}

\usepackage{graphicx}
\usepackage{dcolumn}
\usepackage{bm}

\usepackage{bbm}

\usepackage{hyperref}
\usepackage{color}

\usepackage{ulem}

\begin{document}


\title{Higher Berry curvature,  second Chern numbers and magnetoelectric coupling in crystalline insulators}%

\author{Niclas Heinsdorf*}
\affiliation{Department of Physics and Institute for Quantum Information and Matter,
California Institute of Technology, Pasadena, CA 91125, USA}
\email{niclas@caltech.edu}

\author{Ken Shiozaki}
\affiliation{Center for Gravitational Physics and Quantum Information, Yukawa Institute for Theoretical Physics, Kyoto University, Kyoto 606-8502, Japan}

\date{\today}

\begin{abstract}
{\color{black}We compute the second Chern number of a four-dimensional lattice Chern insulator in a manifestly quantized manner using the higher three-form Berry curvature of a family of one-dimensional ground states represented by infinite matrix product states (iMPS).} We calculate the topological phase diagram of the associated Dixmier--Douady--Kapustin--Spodyneiko (DDKS) number as a function of the model's mass term, and show that it is exactly congruent to the phase diagram in terms of the second Chern number, the analytic expression of which is known for this particular model. This agreement demonstrates that higher Berry curvature can be used to compute second Chern numbers in a manifestly quantized manner. Motivated by the connection between the second Chern form and the Chern--Simons axion coupling, we study magnetoelectric coupling in three dimensions and its relation to higher Berry phases. {\color{black}We find agreement between higher Berry phases and Chern–Simons axion angles in two models, which supports higher Berry geometry as a promising framework for describing magnetoelectric coupling.}
\end{abstract}

\maketitle


\section{\label{sec:intro}Introduction}
Topological invariants provide a powerful way to organize families of quantum many-body ground states. A familiar example is the use of Berry curvature to characterize parameterized families of finite-dimensional ground states. In recent years, this perspective has been extended beyond zero-dimensional systems to families of spatially extended short-range entangled states, where new geometric structures and higher-degree invariants appear{\color{black}\cite{Kit11,Kit13,Kit15,KapustinSpodyneiko_Berry_curvature20}}.

Kitaev conjectured that the topological classification of short-range entangled invertible phases admits a generalization to families of ground states with infinite spatial extent in arbitrary dimension\cite{Kit11,Kit13,Kit15}. Here the dimension in question is the physical dimension of the quantum state, rather than the dimension of the parameter space. This distinction is important. In the Haldane model\cite{haldanemodel}, for instance, the Chern number characterizes a family of periodic Bloch states over the two-dimensional Brillouin zone. Since the Hamiltonian is diagonal in momentum, momentum plays the role of a parameter. One is therefore effectively considering a family of zero-dimensional two-level systems parametrized by points on the two-torus $T^2$, rather than a family of spatially extended two-dimensional many-body states. In this setting, the first Chern number is obtained by integrating the Berry curvature over the Brillouin zone. Numerically, this is often implemented by discretizing the Brillouin zone and evaluating Berry phases\cite{Pancharatnam1956,berry1984quantal} around the simplices of a triangulation or mesh. Summing these contributions gives the Chern number, as in the Fukui-Hatsugai-Suzuki algorithm\cite{fukui_hatsugai}. A key feature of this construction is that the {\color{black}output of the} algorithm is \textit{manifestly quantized}, meaning that it returns integer values by construction.

Kapustin and Spodyneiko proposed a higher-dimensional analogue of this Berry-curvature framework\cite{KapustinSpodyneiko_Berry_curvature20}. For one-dimensional systems, the ordinary Berry curvature, a two-form, is replaced by a three-form; for two-dimensional systems, it is replaced by a four-form; and similarly in higher spatial dimensions. Correspondingly, the relevant parameter spaces must have sufficiently high dimension to support such higher differential forms. These higher Berry curvatures may be viewed as probes of degeneracies or gapless loci in the phase diagram of a family of Hamiltonians. Just as integrating the ordinary Berry curvature produces the Chern number, integrating the one-dimensional higher Berry curvature yields a quantized higher invariant $\nu$ --- referred to as the Dixmier--Douady--Kapustin--Spodyneiko (DDKS) number. Until recently, explicit computations of the higher Berry curvature were limited mainly to certain free-fermion systems\cite{KapustinSpodyneiko_Berry_curvature20} and to spin models that reduce to an effective two-site problem\cite{flow}. Related developments include Refs.~\cite{Kapustin_Sopenko_Local_Noether_22,Artymowicz_Kapustin_Sopenko_Quantization_23,Bachmann2024}. 

{\color{black}Alongside these developments, higher-form Berry curvatures and gerbe structures have been explored in topological band theory through tensor Berry connections and tensor monopoles\cite{refcomment1_1,refcomment1_2}, including experimental measurements of the associated Dixmier--Douady invariant\cite{refcomment1_3}. Connections to observable responses have also been established: Ref.~\cite{refcomment1_4} related higher Berry fluxes and Dixmier--Douady invariants to magnetoelectric coupling in three-dimensional topological insulators, while Refs.~\cite{refcomment1_4,refcomment1_5} investigated their manifestations in optical responses. These works provide important precedents for the interplay between higher Berry geometry and physical responses studied here, although their definitions of higher Berry curvature and the associated invariants differ from the DDKS construction adopted in the present work.} 

More recently, algorithms have been introduced for computing the higher Berry curvature in generic translationally invariant one-dimensional Hamiltonians\cite{shiozaki2023higher,sommer1}. In these approaches, infinite matrix product states (iMPS)~\cite{SCHOLLWOCK201196} replace Bloch states as the fibers. A key insight is that families of iMPS carry a gerbe structure; see Refs.~\cite{geometry_mps_frank,ohyama2023discretehigherberryphases,gerbe,ohyama2024higherberryconnectionmatrix} for detailed discussions.{\color{black}A gerbe is a higher analogue of a line bundle. When equipped with connection data, it has a three-form curvature, generalizing the two-form curvature familiar from the ordinary Berry connection~\cite{nakahara2018geometry,murray2008introductionbundlegerbes}}. This gerbe structure implies that families of iMPS organize into bundles whose topology is classified by appropriate cohomology classes. In recent work, the corresponding classifying space was constructed explicitly: the space of translationally invariant injective MPS was shown to have weak homotopy type $K(\mathbb{Z},2)\times K(\mathbb{Z},3)$. {\color{black}Here, $K(\mathbb{Z},n)$ denotes a space whose only nonzero homotopy group is $\pi_n=\mathbb{Z}$. In practical terms, families parametrized by an $n$-sphere carry an integer label that is unchanged under continuous deformations. The $K(\mathbb{Z},2)$ factor captures the first Chern number per unit cell, while the $K(\mathbb{Z},3)$ factor captures the higher topology associated with the DDKS number. A nonzero invariant means that the family cannot be continuously deformed into a parameter-independent state while remaining within the space of translationally invariant injective MPS ~\cite{beaudry2025classifying}.}

In this work, we use the algorithm proposed in Ref.~\cite{shiozaki2023higher}, which is a generalization of the Fukui-Hatsugai-Suzuki method\cite{fukui_hatsugai} and provide a discrete formula that yields a manifestly quantized second Chern number for four-dimensional free-fermion systems with lattice translational symmetry. First, let us briefly review prior numerical methods for computing manifestly quantized second Chern numbers. One approach employs the lattice index theorem; in the overlap fermion formulation~\cite{NEUBERGER1998141}, the number of chiral zero modes of the four-dimensional Dirac operator exactly matches the second Chern number of the non-Abelian connection defined on the links~\cite{HASENFRATZ1998125}. To apply the lattice index theorem for the second Chern number over the discretized parameter space, one introduces a virtual four-component Dirac fermion on the vertices coupled with the non-Abelian Berry connection on the links. The overlap Dirac fermion is then constructed via the diagonalization of the Dirac operator to compute the chiral zero modes. For recent developments in lattice index theorems, including $K$-theoretic formulations, see Ref.~\cite{aoki2025ktheoretic} and references therein. 
A more direct approach in the real-space formalism exploits the fact that the second Chern number corresponds to the four-dimensional quantum Hall coefficient~\cite{qi_topological_field_theory}. In this method, the second Chern number is extracted from the magnetic field dependence of the occupied state count by applying external magnetic fields to the $xy$ and $yz$ planes of the model on the original four-dimensional real-space lattice~\cite{second_chern_number_magnetic_field}.
Note that both methods require computing the spectrum of Dirac fermions on a four-dimensional lattice torus where translational symmetry is broken by external fields.

The remainder of the paper is organized as follows: In Sec.~\ref{sec:4d_chern} we review the field theory and lattice model of the four-dimensional Chern insulator that we then rewrite as a collection of one-dimensional Hamiltonians defined over three-dimensional parameter space. In Sec.~\ref{sec:higher_berry} we proceed to explicitly compute the higher Berry curvature and the related topological invariants using the above-mentioned algorithm. In Sec.~\ref{sec:2nd_chern} we compute second Chern numbers of the same model and compare it to the phase diagram generated by the DDKS number. In Sec.~\ref{sec:mec} we review the connection between the second Chern number and the Chern-Simons axion angle $\theta_{\rm CS}$ in the context of orbital magnetoelectric coupling. We examine numerical values of $\theta_{\rm{CS}}$ for the four-dimensional Dirac model as well as the Hopf insulator, which we contrast with $\theta_{\rm{HBP}}$ --- a purely geometric quantity derived from higher Berry phases. Lastly, we summarize our results and address open questions in Sec.~\ref{sec:conclusion}. 
\begin{figure}
    \centering
    \includegraphics[width=0.8\linewidth]{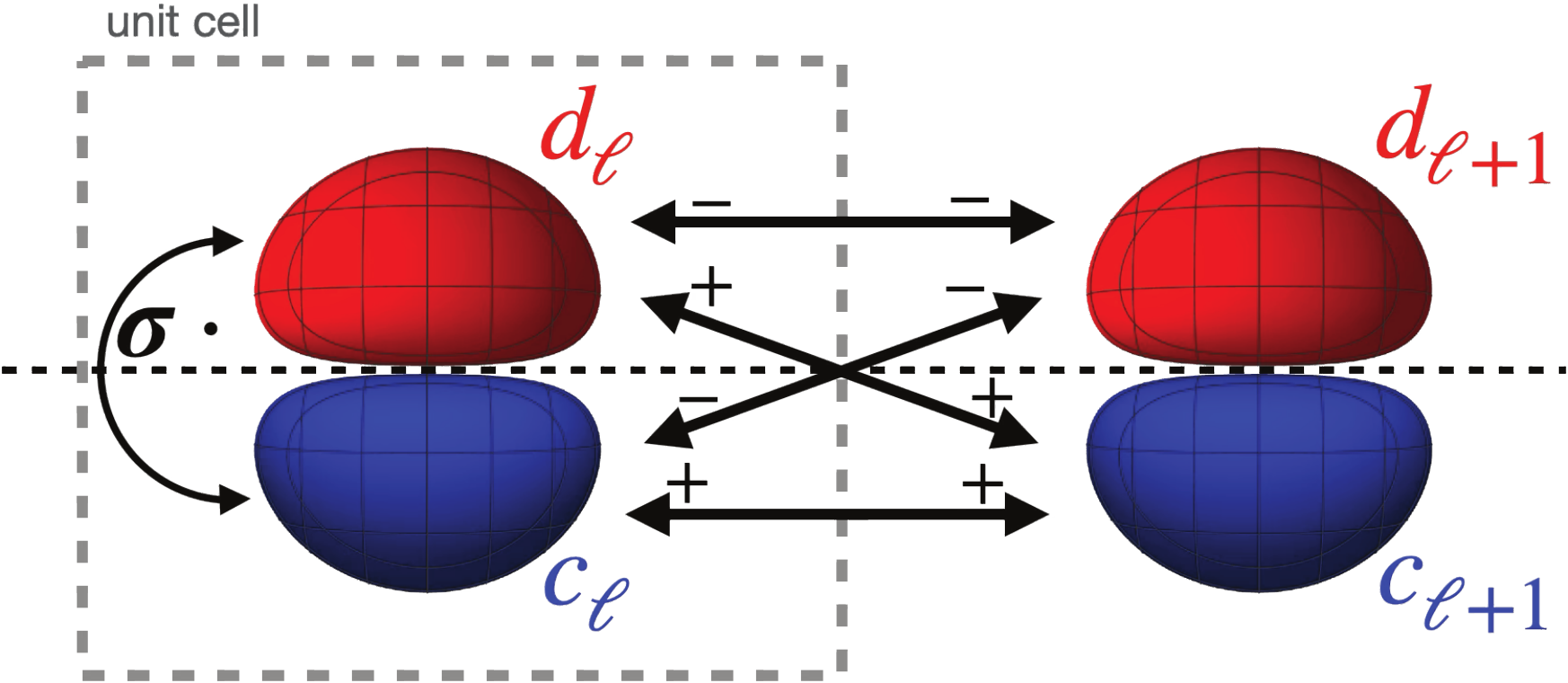}
    \caption{Schematic of the one-dimensional model with two orbitals $c$ (blue) and $d$ (red) per unit cell labeled by $\ell$. The $\boldsymbol{\lambda}$-dependent onsite {\color{black}spin-orbit coupling} (SOC) is indicated by $\boldsymbol{\sigma}$ and the black arrows indicate the phases of the intersite hopping terms. Not shown in this schematic is the mass term which depends on both $\boldsymbol{\lambda}$ and $m$ \mbox{(see Eq.~\eqref{eq:1D_hamiltonian})}. Reproduced from Ref.~\cite{Heinsdorf_2025}.}
    \label{fig:fig1}
\end{figure}

\section{\label{sec:4d_chern}Four-Dimensional Chern Insulator}

We begin from the $(4+1)$-dimensional Dirac Hamiltonian\cite{dirac_4p1_golterman,dirac_4p1_creutz} 
\begin{align}
    \mathcal{H} = \int d^4x\left[\psi^\dagger(x)\Gamma^\mu(-i\partial_\mu)\psi(x) + m\psi^\dagger \Gamma^0\psi \right],
\end{align}
where $\mu$ runs over four spatial dimensions. $\Gamma^0$ and $\Gamma^\mu$ are the five anti-commuting Dirac matrices. 

The lattice version of this model is given by\cite{qi_topological_field_theory}
\begin{align}\label{eq:tb_4d_hamiltonian}
    \mathcal{H}_\mathbf{k} = \psi_\mathbf{k}^\dagger\left[ \sum_\mu \sin k_\mu \Gamma^\mu + \left(m + c\sum_\mu \cos k_\mu \right)\Gamma^0 \right]\psi_\mathbf{k},
\end{align}
with momenta on the four-torus $\mathbf{k}\in T^4$. {\color{black}Here, $c$ is a free parameter of the lattice regularization that controls the strength of the momentum-dependent mass term.} From now on, we set $c=1$. The spectrum of this model is 
\begin{align}
    E_\pm(\mathbf{k}) = \pm \sqrt{\left(m + \sum_\mu \cos k_\mu \right)^2 + \sum_\mu \sin^2 k_\mu}.
\end{align}
We can characterize the four-dimensional Dirac Hamiltonian in terms of the second Chern number $\mathrm{ch}_2$. The topological phase diagram can be derived analytically and is given by
\begin{align}\label{eq:ch2_phase_diagram}
    \mathrm{ch}_2(m)=\left\{\begin{array}{ll}0,\quad &|m| > 4\\ 1,\quad &-4 < m < -2\\-3,\quad &-2 < m < 0 \\3,\quad &0 < m < 2\\-1,\quad &2 < m < 4\end{array}\right.
\end{align}
Finally, we observe that the four-dimensional BZ may be decomposed as
\begin{align}\label{eq:four_torus_product}
    T^4 =  S^1 \times T^3.
\end{align}
This product structure will be used in \mbox{Sec.~\ref{sec:2nd_chern}}, where the K\"unneth formula (see Eq.~\eqref{eq:kuenneth}) allows us to relate the DDKS number $\nu$ of a family of iMPS parametrized by $T^3$ to the second Chern number $\mathrm{ch}_2$. 

For the moment, we use the decomposition in Eq.~\eqref{eq:four_torus_product} to reinterpret the four-dimensional tight-binding Hamiltonian of Eq.~\eqref{eq:tb_4d_hamiltonian} as a three-parameter family of one-dimensional Hamiltonians. More precisely, we Fourier transform only the momentum component $k_4$, thereby obtaining Hamiltonians $\{\mathcal{H}_{\boldsymbol{\lambda}} \}$ that are spatially extended in one direction and depend on the parameter $\boldsymbol{\lambda}=(\lambda_1, \lambda_2, \lambda_3)^\top$. The result is
\begin{widetext}
    \begin{equation}\label{eq:1D_hamiltonian}
        \mathcal{H}_{\boldsymbol{\lambda}} = \sum_\ell \left[\mathbf{c}^\dagger_{\ell+1}\mathbf{c}_{\ell} - \mathbf{d}^\dagger_{\ell+1}\mathbf{d}_{\ell} + \mathbf{d}^\dagger_{\ell+1}\mathbf{c}_{\ell} - \mathbf{c}^\dagger_{\ell+1}\mathbf{d}_{\ell} + \sum_{n=1}^3 \sin \lambda_n \mathbf{c}^\dagger_\ell \sigma_n \mathbf{d}_\ell + \frac{1}{2} \left(\sum_{n=1}^3 \cos \lambda_n + m\right)\left(\mathbf{c}^\dagger_{\ell}\mathbf{c}_{\ell} - \mathbf{d}^\dagger_{\ell}\mathbf{d}_{\ell}\right)\right] + \mathrm{h.c.}
    \end{equation}
\end{widetext}
The parameters $\boldsymbol{\lambda}$ are the three momentum components of $\mathbf{k}$ that remain after Fourier transforming the fourth direction, relabeled for notational clarity. Eq.~\eqref{eq:1D_hamiltonian} therefore defines a family of translationally invariant one-dimensional chains with two orbitals in each unit cell. The unit-cell index is denoted by $\ell$, and the two orbitals are created by $c_{\ell s}^\dagger$ and $d_{\ell s}^\dagger$, respectively. A schematic representation of the model is shown in \mbox{Fig.~\ref{fig:fig1}}.

The fermionic operators $c_{\ell s}$ and $d_{\ell s}$ obey the canonical anticommutation relations. After a Jordan-Wigner transformation, the Hamiltonian can be represented by a standard matrix product operator (MPO)~\cite{tenpy1}. The spectrum closes only at the critical values $m=0, \pm2, \pm4$. Away from these points, the system is gapped. Its ground state is then a translationally invariant one-dimensional invertible state and can be approximated, to arbitrary precision, by an iMPS with finite bond dimension\cite{SCHOLLWOCK201196}. {\color{black}For fixed noncritical $m$, the Hamiltonian $\mathcal{H}_{\bm{\lambda}}$ remains gapped throughout the parameter space $T^3$.} Assuming that the corresponding ground states have trivial short-range entanglement, we may choose the associated MPS representatives to be injective\cite{MPS}. These properties provide the starting point for the computation of the higher Berry curvature and the associated topological invariant in the next section.

\section{\label{sec:higher_berry}Higher Berry Curvature}

\begin{figure}
    \centering
    \includegraphics[width=1\linewidth]{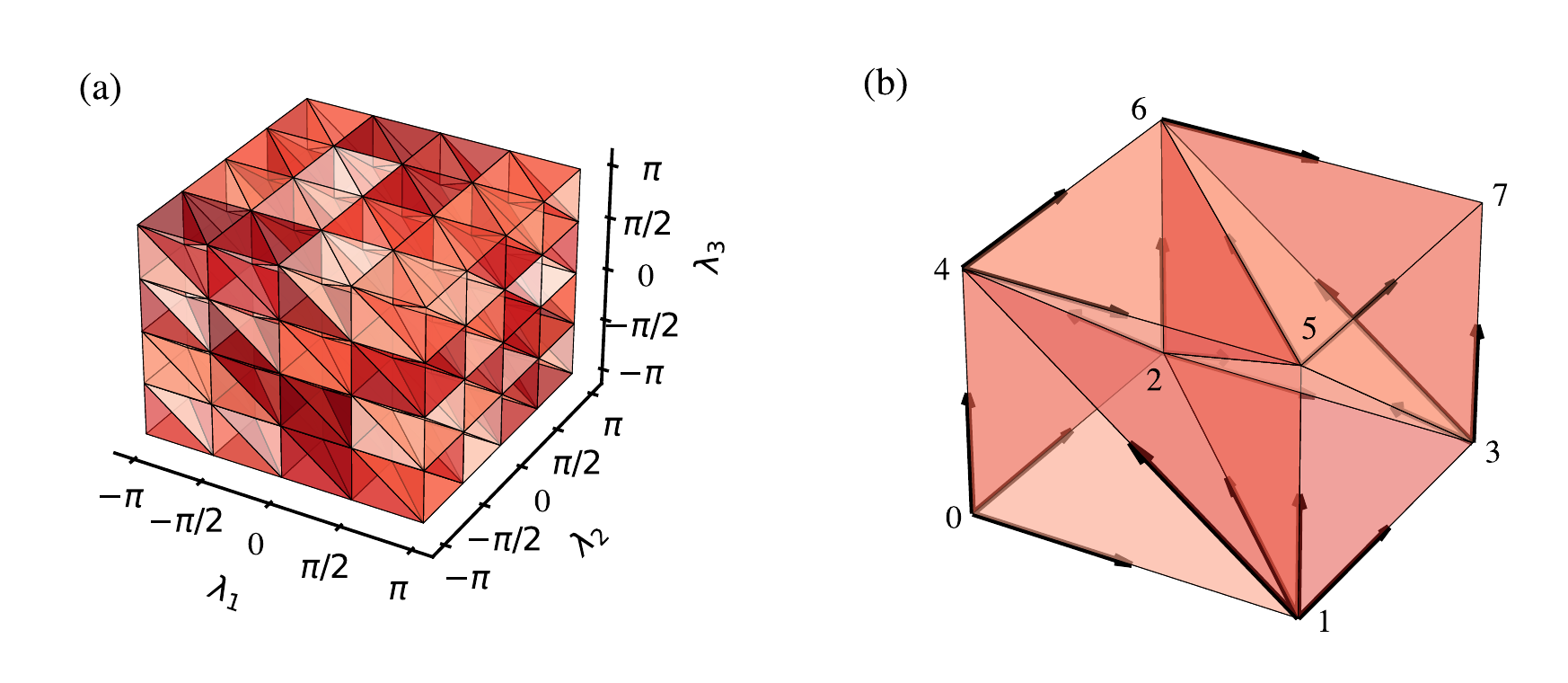}
    \caption{(a) Triangulation of the three-dimensional parameter space $|T^3|$ spanned by $\boldsymbol{\lambda}$ on a $(N+1)^3$-grid. (b) Triangulation of a single cube with eight labeled vertices into six tetrahedra $\Delta^3$. Each tetrahedron has four triangular faces $\Delta^2$. The arrows attached to the vertices indicate the orientation of the corresponding tetrahedron. Reproduced from Ref.~\cite{Heinsdorf_2025}.}
    \label{fig:fig2}
\end{figure}

Following Ref.~\cite{shiozaki2023higher}, we now proceed to review the construction of the higher Berry curvature, before we present our numerical results. For a more detailed derivation, we refer to the original publication.

We now describe the discrete construction used to evaluate the higher Berry curvature for a family of one-dimensional ground states over the parameter space $T^3$. We begin by introducing a uniform grid with $(N+1)^3$ vertices covering the three-torus parametrized by $\boldsymbol{\lambda}$. A triangulation of the parameter space $|T^3|$ is shown in Fig.~\ref{fig:fig2}(a). At each vertex $\boldsymbol{\lambda}_n$ of this grid, we use iDMRG as implemented in the TeNPy package~\cite{tenpy1,tenpyv1} to obtain the corresponding ground-state iMPS, denoted by $A_{\boldsymbol{\lambda}_n}^i$. Each iMPS tensor consists of $d$ complex $D\times D$ matrices. Here $d$ is the local Hilbert-space dimension, or physical leg, and is fixed by the local MPO representation, while $D$ is the bond dimension, or virtual leg. Since the required bond dimension depends on the local entanglement structure and on the size of the gap, $D$ may vary from one grid point to another.

For a chain of $L$ unit cells at fixed $\boldsymbol{\lambda}$, the corresponding finite-size MPS is written as
\begin{align}
|\{A^i\}_i\rangle_L := \sum_{i_1,\dots,i_L}
\mathrm{Tr}[A^{i_1}\cdots A^{i_L}] |i_1\cdots i_L\rangle.
\label{eq:mps_Lsites}
\end{align}
We call an iMPS (left-)canonical, if it satisfies 
\begin{align}
\sum_{i=1}^d A^i A^{i\dag} = \mathbbm{1}_{D\times D},
\label{eq:right_cano}
\end{align}
with a positive definite diagonal matrix $\Lambda^2$,
\begin{align}
\mathrm{Tr}[\Lambda^2]=1, \label{eq:normalization_Lambda}
\end{align}
such that
\begin{align}
\sum_{i=1}^d A^{i\dag} \Lambda^2 A^i = \Lambda^2
\label{eq:left_cano}
\end{align}
holds. $\Lambda$ is a matrix with the Schmidt eigenvalues as diagonal entries~\cite{tenpy1}.

Now, consider two iMPS tensors ${A^i_0}$ and ${A^i_1}$, where the labels $0$ and $1$ simply distinguish the two states and do not refer to specific coordinate directions in parameter space. Their mixed transfer matrix is the linear map\cite{MPS,Perez-Garcia_Wolf_Sanz_Verstraete_Cirac_String_Order08}
\begin{align}
    T_{01}(X):=\sum_{i=1}^d A^i_0 X A^{i\dagger}_1. 
    \label{eq:mixed_TM_01}
\end{align}
The transition function associated with this pair of tensors is defined by the dominant eigenvector $V_{01}$ of $T_{01}$ with eigenvalue $\eta_{01}$:
\begin{align}
    T_{01}(V_{01})=
    \sum_{i=1}^d A_0^i V_{01} A_1^{i\dagger}
    =\eta_{01} V_{01}\label{eq:V01_def}.
\end{align}
Taking the Hermitian conjugate of Eq.~\eqref{eq:V01_def} gives $\eta^*_{01}V_{01}^\dagger$ and therefore, up to an overall constant,
\begin{align}\label{eq:V_01_hc}
    V_{01} = V^\dagger_{10}.
\end{align}
When the two tensors are injective and correspond to nearby points in parameter space, the leading eigenvalue $\eta_{01}$ is nondegenerate and separated from the rest of the spectrum of $T_{01}$ by a finite gap. 

The construction must remove the gauge freedom inherent in the MPS representation, in close analogy with how the Fukui-Hatsugai-Suzuki algorithm cancels the $U(1)$ gauge ambiguity of Bloch states\cite{fukui_hatsugai}. For fixed bond dimension $D$, the MPS gauge group is the projective unitary group $PU(D) = U(D)/U(1)$\cite{MPS} defined by the action
\begin{align}\label{eq:gauge_mps}
    A^i_n \mapsto e^{i\theta_n} W_n A^i_n W^\dagger_n,
\end{align}
where the gauge transformation may vary over parameter space. Under this transformation, the Schmidt matrix changes as
\begin{align}\label{eq:gauge_trafo_lambda}
    \Lambda_n \mapsto W_n\Lambda_nW_n^\dagger
\end{align}
and the transition function transforms according to
\begin{align}
    V_{01} \mapsto W_0 V_{01} W_1^\dagger.\label{eq:gauge_trafo_V01}
\end{align}
In addition, because $V_{01}$ is obtained as an eigenvector of the mixed transfer matrix, it is only fixed up to a nonzero complex scalar. Thus there is a second ambiguity
\begin{align}
    V_{01} \mapsto z_{01} V_{01} .\label{eq:gauge_trafo_V01_nom2}
\end{align}
A discrete higher Berry curvature must be invariant under both types of transformations.

For an oriented triangular loop $\Delta^2 = (\boldsymbol{\lambda}_0\boldsymbol{\lambda}_1\boldsymbol{\lambda}_2)$ in parameter space, we define a phase
\begin{align}
\phi(\Delta^2) = \mathrm{Arg}\ \mathrm{Tr} [\Lambda_{\boldsymbol{\lambda}_0}^\frac{2}{3} V_{\boldsymbol{\lambda}_0\boldsymbol{\lambda}_1} \Lambda_{\boldsymbol{\lambda}_1}^\frac{2}{3} V_{\boldsymbol{\lambda}_1\boldsymbol{\lambda}_2} \Lambda_{\boldsymbol{\lambda}_2}^\frac{2}{3} V_{\boldsymbol{\lambda}_2\boldsymbol{\lambda}_0}].
\label{eq:dis_Berry_connection}
\end{align}
Here, the vertices $\boldsymbol{\lambda}_n$ are those of the triangle $\Delta^2$, as illustrated in Fig.~\ref{fig:fig2}(b). The factors of $\Lambda$ weight the transition functions by the Schmidt spectrum, which allows the expression to accommodate bond dimensions that vary along the loop. Applying the gauge transformations in Eq.~\eqref{eq:gauge_trafo_lambda} and Eq.~\eqref{eq:gauge_trafo_V01} gives
\begin{align}
\phi(\Delta^2) = \mathrm{Arg}\ \mathrm{Tr} [&W_{\boldsymbol{\lambda}_0}\Lambda_{\boldsymbol{\lambda}_0}^\frac{2}{3}W_{\boldsymbol{\lambda}_0}^\dagger W_{\boldsymbol{\lambda}_0} V_{\boldsymbol{\lambda}_0\boldsymbol{\lambda}_1}W_{\boldsymbol{\lambda}_1}^\dagger \nonumber \\\times &W_{\boldsymbol{\lambda}_1}\Lambda_{\boldsymbol{\lambda}_1}^\frac{2}{3} W_{\boldsymbol{\lambda}_1}^\dagger W_{\boldsymbol{\lambda}_1}V_{\boldsymbol{\lambda}_1\boldsymbol{\lambda}_2} W^\dagger_{\boldsymbol{\lambda}_2}\nonumber \\ 
\times&W_{\boldsymbol{\lambda}_2}\Lambda_{\boldsymbol{\lambda}_2}^\frac{2}{3} W_{\boldsymbol{\lambda}_2}^\dagger W_{\boldsymbol{\lambda}_2}V_{\boldsymbol{\lambda}_2\boldsymbol{\lambda}_0}W_{\boldsymbol{\lambda}_0}^\dagger].
\end{align}
Because the unitary matrices cancel pairwise, $\phi(\Delta^2)$ is invariant under the $PU(D)$ gauge action. It is not, however, invariant under the scaling ambiguity in Eq.~\eqref{eq:gauge_trafo_V01_nom2}. Extracting the factors $z_{nm}$ from the trace and using additivity of the argument modulo $2\pi$, one obtains
\begin{align}\label{eq:z_factor_ambiguity}
    \phi(\Delta^2) \mapsto \phi(\Delta^2) + \mathrm{Arg}(z_{\boldsymbol{\lambda}_0\boldsymbol{\lambda}_1}) + \mathrm{Arg}(z_{\boldsymbol{\lambda}_1\boldsymbol{\lambda}_2}) + \mathrm{Arg}(z_{\boldsymbol{\lambda}_2\boldsymbol{\lambda}_0}).
\end{align}

This residual freedom reflects the fact that, for one-dimensional systems, the higher Berry curvature is a three-form\cite{KapustinSpodyneiko_Berry_curvature20}. The relevant local object is therefore not a flux through a two-dimensional face, but a contribution associated with a three-dimensional volume element. Gauge invariance under Eq.~\eqref{eq:gauge_trafo_V01_nom2} is recovered by combining the phases on the closed two-dimensional boundary of such a volume element.

As shown in \mbox{Fig.~\ref{fig:fig2}(b)}, a small cube in the parameter-space grid, with vertices $\boldsymbol{\lambda}_n$ for $n=0,1,\ldots,7$, is decomposed into six tetrahedra $\Delta^3$. For an oriented tetrahedron, the boundary is
\begin{align}
    \partial \Delta^3 = (\boldsymbol{\lambda}_1\boldsymbol{\lambda}_2\boldsymbol{\lambda}_3) - (\boldsymbol{\lambda}_0\boldsymbol{\lambda}_2\boldsymbol{\lambda}_3) + (\boldsymbol{\lambda}_0\boldsymbol{\lambda}_1\boldsymbol{\lambda}_3) - (\boldsymbol{\lambda}_0\boldsymbol{\lambda}_1\boldsymbol{\lambda}_2),
\end{align}
with the orientations of the triangular faces indicated by the arrows in the figure. This is the higher-dimensional analogue of the usual lattice Berry-curvature construction: for zero-dimensional states, Stokes' theorem relates the Berry phase around the boundary of a small plaquette or simplex to the Berry-curvature flux through that two-dimensional element.

\begin{figure}
    \centering
    \includegraphics[width=1\linewidth]{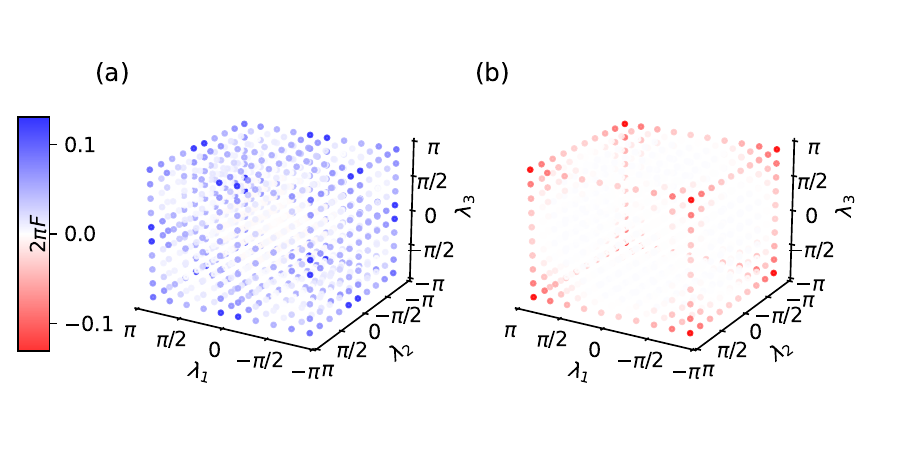}
    \caption{Higher Berry curvature $2\pi F$ over discretized parameter space for (a) $m=1$ and (b) $m=3$ with DDKS numbers $\nu = 3$ and $\nu=-1$ respectively. Compare to Eq.~\eqref{eq:ch2_phase_diagram}. The absolute value of $2\pi F$ is largest where the energy gap is smallest. Reproduced from Ref.~\cite{Heinsdorf_2025}.}
    \label{fig:fig3}
\end{figure}

In the present setting, the higher Berry curvature assigned to a tetrahedron $\Delta^3$ is approximated by the oriented sum of the phases on its boundary faces:
\begin{align}
    F(\Delta^3) = &\phi(\boldsymbol{\lambda}_1\boldsymbol{\lambda}_2\boldsymbol{\lambda}_3)-\phi(\boldsymbol{\lambda}_0\boldsymbol{\lambda}_2\boldsymbol{\lambda}_3) \\+&\phi(\boldsymbol{\lambda}_0\boldsymbol{\lambda}_1\boldsymbol{\lambda}_3)-\phi(\boldsymbol{\lambda}_0\boldsymbol{\lambda}_1\boldsymbol{\lambda}_2).
\end{align}
This combination is invariant under the scalar ambiguity of Eq.~\eqref{eq:gauge_trafo_V01_nom2}. Indeed, Eq.~\eqref{eq:V_01_hc} implies $z_{nm} = z^*_{mn}$, and hence $\mathrm{Arg}(z^*_{nm}) = -\mathrm{Arg}(z_{nm})$ and we can write 
\begin{align}
    F(\Delta^3) \mapsto F(\Delta^3) &+ \mathrm{Arg}(z_{\boldsymbol{\lambda}_1\boldsymbol{\lambda}_2}) + \mathrm{Arg}(z_{\boldsymbol{\lambda}_2\boldsymbol{\lambda}_3}) + \mathrm{Arg}(z_{\boldsymbol{\lambda}_3\boldsymbol{\lambda}_1}) \nonumber \\
    &- \mathrm{Arg}(z_{\boldsymbol{\lambda}_0\boldsymbol{\lambda}_2}) - \mathrm{Arg}(z_{\boldsymbol{\lambda}_2\boldsymbol{\lambda}_3}) - \mathrm{Arg}(z_{\boldsymbol{\lambda}_3\boldsymbol{\lambda}_0}) \nonumber \\
    &+ \mathrm{Arg}(z_{\boldsymbol{\lambda}_0\boldsymbol{\lambda}_1}) + \mathrm{Arg}(z_{\boldsymbol{\lambda}_1\boldsymbol{\lambda}_3}) + \mathrm{Arg}(z_{\boldsymbol{\lambda}_3\boldsymbol{\lambda}_0}) \nonumber \\
    &- \mathrm{Arg}(z_{\boldsymbol{\lambda}_0\boldsymbol{\lambda}_1}) - \mathrm{Arg}(z_{\boldsymbol{\lambda}_1\boldsymbol{\lambda}_2}) - \mathrm{Arg}(z_{\boldsymbol{\lambda}_2\boldsymbol{\lambda}_0}) \nonumber \\
    &=F(\Delta^3),
\end{align}
where all residual phase factors canceled. 

Finally, summing $F(\Delta^3)$ over all oriented tetrahedra in a closed three-dimensional triangulation gives a manifestly quantized invariant. For the family of Hamiltonians in Eq.~\eqref{eq:1D_hamiltonian} over $T^3$, we therefore compute
\begin{align}\label{eq:integral_higherF_eq_nu}
    \nu = \frac{1}{2\pi}\sum_{\Delta^3\in |T^3|}F(\Delta^3).
\end{align}
The resulting integer $\nu$ represents the corresponding class in the integral third cohomology of the three-torus, $H^3(T^3, \mathbbm{Z})$.

In Fig.~\ref{fig:fig3} we show the distribution of higher Berry curvature across the $\boldsymbol{\lambda}$-grid with the resulting topological invariants $\nu$ for different values of $m$. The Berry curvature is largest where the energy gap is small. Compare to Ref.~\cite{qi_topological_field_theory}. We find the phase diagram to be exactly congruent with that given in terms of the second Chern number from Eq.~\eqref{eq:ch2_phase_diagram}, which we discuss in the following section.

\section{\label{sec:2nd_chern}Second Chern Number}

\begin{figure}
    \centering
    \includegraphics[width=0.95\linewidth]{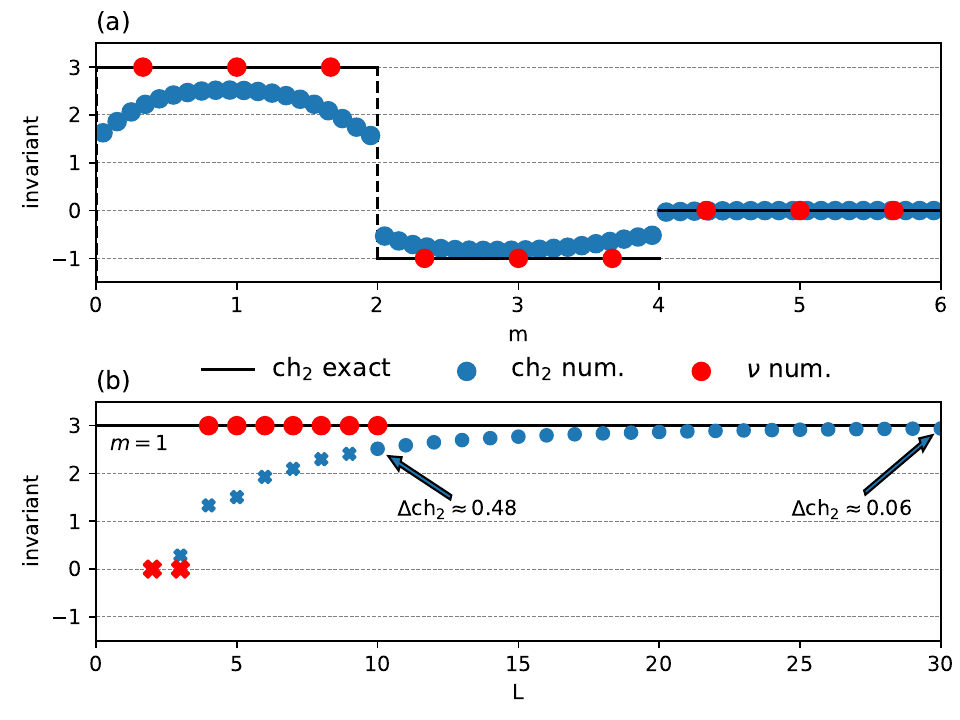}
    \caption{(a) Topological phase diagram of the four-dimensional Chern insulator as a function of $m$. The second Chern number $\mathrm{ch}_2$ of the four-dimensional lattice model given in Eq.~\eqref{eq:tb_4d_hamiltonian} as obtained from the analytical formula in Ref.~\cite{qi_topological_field_theory} (black) and using the algorithm {\color{black} without manifestly quantized output} proposed in Ref.~\cite{algo_2nd_Chern} (blue). The numerical results for the DDKS number $\nu$ of the Hamiltonian in Eq.~\eqref{eq:1D_hamiltonian} using the algorithm {\color{black}with manifestly quantized output} from Ref.~\cite{shiozaki2023higher} (red) match the phase diagram of $\mathrm{ch}_2$ perfectly. (b) Numerical results for $\mathrm{ch}_2$ (blue) and $\nu$ (red) for $m=1$ as a function of {\color{black} the linear grid resolution $L$, using $L^4$ momentum-space points and $L^3$ parameter-space points, respectively.} The black line indicates the exact value of $\mathrm{ch}_2=\nu=3$. The sum of higher Berry curvature yields the correct invariant down to {\color{black}$L=4$}. The error of the not manifestly quantized second Chern number for two data points is given explicitly (indicated by the arrows). {\color{black}The legend between the two panels applies to both panels (a) and (b).} Reproduced from Ref.~\cite{Heinsdorf_2025}.}
    \label{fig:fig4}
\end{figure}

The second Chern number is defined as\cite{qi_topological_field_theory}
\begin{align}
    \mathrm{ch}_2 = \frac{1}{32\pi^2}\int_{\mathrm{BZ}}\mathrm{Tr}\left[F \wedge F\right],
\end{align}
where the trace is taken over the occupied states at fixed four-momentum $\mathbf{k}$. 
Physically, $\mathrm{ch}_2$ appears as the coefficient of the electromagnetic Chern-Simons effective action for a four-dimensional generalization of the Chern insulator~\cite{4dchernscience}. It therefore determines the corresponding higher-dimensional Hall response. Equivalently, this response can be expressed as the change in the many-body first Chern number in the $xy$ plane, defined from the ground-state wave function, upon applying a unit magnetic flux through the $zw$ plane~\cite{second_chern_number_magnetic_field}. For noninteracting lattice models, the second Chern number can be evaluated numerically using lattice gauge theory methods. In general, such discretizations {\color{black} do not have manifestly quantized output} at finite mesh size. Instead, the computed value converges to an integer in the continuum limit, with an error controlled by the momentum-space lattice spacing $\Delta k$~\cite{algo_2nd_Chern}. This convergence can be improved by using adaptive meshes~\cite{more_algo3}. By contrast, the higher-Berry-curvature construction discussed above yields a manifestly quantized integer after summing over the discretized three-dimensional parameter space. This distinction makes it useful to compare the two quantities in the present model, not as a benchmark of numerical efficiency, but as a check of the proposed correspondence between $\nu$ and $\mathrm{ch}_2$.

We apply the algorithm of Ref.~\cite{algo_2nd_Chern} to the lattice four-dimensional Chern insulator defined in Eq.~\eqref{eq:tb_4d_hamiltonian}. The calculation is performed on a $10^4$ momentum-space grid. The grid is chosen so that the values of $k_2$, $k_3$, and $k_4$ are located at the centers of the three-dimensional hypercubes used in the higher-Berry-curvature calculation of the preceding section. This choice allows the two discretizations to be compared on closely aligned parameter-space meshes.

Fig.~\ref{fig:fig4}(a) shows the exact phase diagram of the second Chern number as a function of the mass parameter $m$, together with the numerical values of $\mathrm{ch}_2$ and $\nu$ obtained from the respective discretized calculations. The two phase diagrams agree throughout the gapped regions. For negative mass, the sign of both invariants is reversed. As expected, the higher-Berry-curvature sum gives an exactly quantized result by construction. The lattice-gauge-theory estimate of $\mathrm{ch}_2$, on the other hand, approaches the integer value only after extrapolation or sufficient mesh refinement. Near a topological phase transition, where the gap becomes small, the finite-grid value of $\mathrm{ch}_2$ can deviate from the limiting integer by more than one.

The convergence with grid size is shown in Fig.~\ref{fig:fig4}(b) for $m=1$. In this case, the higher-Berry-curvature calculation already gives the correct integer on a $4^3$ grid. The lattice-gauge-theory calculation of $\mathrm{ch}_2$ requires a four-dimensional grid of approximately $10^4$ points before the deviation satisfies $\Delta \mathrm{ch}_2<0.5$. For a $30^4$ grid, the error is reduced to approximately $\Delta \mathrm{ch}_2\simeq 0.06$. We note that $m=1$ is a comparatively favorable value for the lattice calculation, since it lies midway between the neighboring critical points at $m=0$ and $m=2$.

The purpose of this comparison is to demonstrate the agreement between the phase diagrams determined by $\mathrm{ch}_2$ and by the DDKS number $\nu$. We do not intend to present the iMPS-based method as a replacement for standard lattice-gauge-theory calculations of the second Chern number in noninteracting systems. The two approaches take different types of input and compute different, although related, topological invariants. In the present free-fermion model, the second Chern number is naturally accessed from the Bloch Hamiltonian, while $\nu$ is obtained from a family of one-dimensional ground states represented as iMPS.

For completeness, we briefly summarize the computational ingredients of the two calculations. In the lattice-gauge-theory approach, the dominant step is the diagonalization of the Bloch Hamiltonian at each point of the four-dimensional grid. For Eq.~\eqref{eq:tb_4d_hamiltonian}, this amounts to diagonalizing $N^4$ Hermitian $4\times4$ matrices, apart from the subsequent construction of link variables and Berry-curvature densities. In the higher-Berry-curvature calculation, one instead computes ground-state iMPS on an $N^3$ parameter-space grid. The convergence of iDMRG depends on the required bond dimension, with a typical scaling of order $\mathcal{O}(D^3)$\cite{tenpy1}. Both procedures are naturally parallel over their respective momentum or parameter grids.

For the model studied here, we find that a bond dimension $D=64$ is sufficient. We use convergence criteria given by a maximal energy error $\Delta E=10^{-8}$ and a maximal entanglement-entropy error $\Delta S=10^{-6}$. After the iMPS ground states have been obtained, the higher-Berry-curvature construction also requires the dominant eigenvector of the mixed transfer matrix on each link. For two iMPS of bond dimension $D$, this transfer matrix has dimension $D^2\times D^2$. These link calculations are again independent and can be parallelized.

Finally, we emphasize that for the specific Hamiltonian in Eq.~\eqref{eq:tb_4d_hamiltonian}, an analytical expression for the Berry-curvature density is available. In general, tensor-network methods such as iDMRG are expected to be most useful in settings where the ground states are interacting and cannot be described directly in terms of single-particle Bloch wave functions. Possible interacting systems with nontrivial DDKS number are discussed in Sec.~\ref{sec:conclusion}. Next, we clarify the relation between $\mathrm{ch}_2$ and $\nu$.


In a four-dimensional Chern insulator, the second Chern number characterizes the integer part of the fourth cohomology $H^4(T^4, \mathbb{Z})$. 
On the other hand, the DDKS number $\nu$ computed in the previous section is an element of the third cohomology $H^3(T^3, \mathbb{Z})$. 
They are related through the K\"unneth formula 
\begin{align}\label{eq:kuenneth}
    H^4(S^1 \times T^3,\mathbb{Z}) &\cong \bigoplus_{p+q=4} H^p(S^1,\mathbb{Z}) \otimes H^q(T^3,\mathbb{Z})\\ &\cong H^1(S^1,\mathbb{Z})\otimes H^3(T^3,\mathbb{Z})  \cong H^3(T^3,\mathbb{Z}),
\end{align}
where the generator of $H^1(S^1,\mathbb{Z})$ corresponds to the additional adiabatic parameter. Under this identification, the second Chern class on $T^4=S^1\times T^3$ is represented by the product of the generator of $H^1(S^1,\mathbb{Z})$ with the three-dimensional class measured by $\nu$; consequently, with a fixed choice of orientation, the corresponding integer invariants agree, $\mathrm{ch}_2=\nu$. {\color{black}For the smooth, gapped, translation-invariant family of local, charge-conserving free-fermion Hamiltonians considered here, the identification of the higher Berry class with the integral of the degree-four Chern character over the momentum circle $S^1$ has been established in Ref.~\cite{KapustinSpodyneiko_Berry_curvature20}.}

To understand how an invariant taking values in $H^3(T^3, \mathbb{Z})$ is constructed from a three-parameter family of one-dimensional free fermions, we employ an analogy with the Thouless pump. 
A Wilson loop can be fundamentally defined for any one-dimensional Hamiltonian independent of external parameters. Applying this definition to a parameterized family of one-dimensional Hamiltonians $H(k_x; \bm{\lambda})$, the occupied Bloch states $U(k_x; \bm{\lambda}) = (u_1(k_x; \bm{\lambda}), \dots, u_n(k_x; \bm{\lambda}))$ at each parameter point $\bm{\lambda}$ constitute the $U(n)$ Wilson loop
\begin{align}
    W(\bm{\lambda}) = \lim_{\mathcal{N} \to \infty} \prod_{l=0,\dots,\mathcal{N}-1}^\leftarrow U\left(\frac{2\pi (l+1)}{\mathcal{N}};\bm{\lambda}\right)^\dagger U\left(\frac{2\pi l}{\mathcal{N}};\bm{\lambda}\right).
\end{align}
Note that this Wilson loop is subject to a $U(n)$ gauge ambiguity at the base point, expressed as $W(\bm{\lambda}) \mapsto V(\bm{\lambda})^\dagger W(\bm{\lambda}) V(\bm{\lambda}), V(\bm{\lambda}) \in U(n)$. 
Assuming the Wilson loop $W(\bm{\lambda})$ can be smoothly defined globally over the parameter space $T^3$, its winding number is given by $\frac{1}{24\pi^2} \int_{T^3} \text{tr} [(W^\dagger d W)^3] \in \mathbb{Z}$, which provides the explicit construction of the element in $H^3(T^3, \mathbb{Z})$. 
However, in general, there exists no method to provide a smooth gauge over the entire parameter space. In contrast, the approach using the DDKS number discussed in this work is independent of such gauge ambiguities.

\section{\label{sec:mec}Orbital Magnetoelectric Coupling}
Since the higher Berry curvature is gauge invariant, it is natural to ask whether it is related to a physical response quantity in three spatial dimensions. We have shown above that the higher Berry curvature can be used to compute the second Chern number, which in four dimensions determines the coefficient of the Chern-Simons effective action. Because the Chern-Simons term in three dimensions is known to enter the orbital magnetoelectric coupling, we investigate its relation to the higher Berry curvature in the following. We start by reviewing orbital magnetoelectricity in crystalline insulators {\color{black}\cite{PhysRevLett.102.146805,PhysRevB.81.205104}}.

\begin{figure}
    \centering
    \includegraphics[width=1\linewidth]{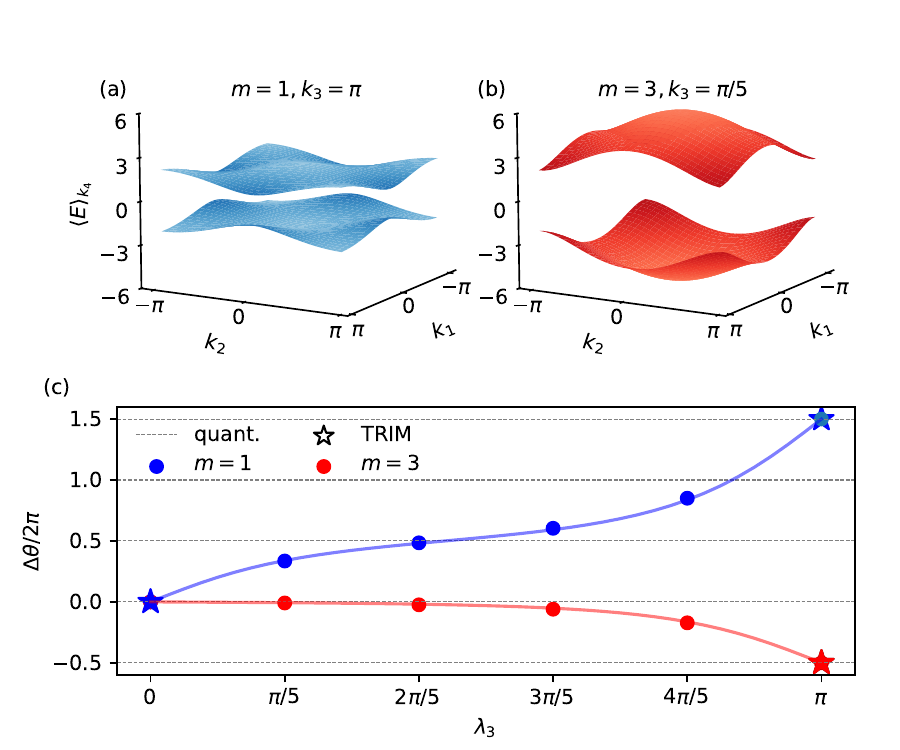}
    \caption{(a), (b) The energy density of Eq.~\eqref{eq:tb_4d_hamiltonian} averaged over $k_4$ at $k_3 = \pi$ with $m=1$ and at $k_3 = \pi/5$ with $m=3$ as a function of $k_1$ and $k_2$, respectively. The spectra are symmetric w.r.t. reflection over the Fermi level. (c) $\Delta\theta_{\rm CS}$ (solid line) and the numerical results for $\Delta\theta_{\rm HBP}$ (scatter plot) as a function of $\lambda_3$ for the Dirac model in Eq.~\eqref{eq:1D_hamiltonian} for $m=1$ (blue) and $m=3$ (red). The stars at $\lambda_3=0$ and $\lambda_3=\pi$ mark the TRIM where $\lambda_3$-slices of Eq.~\eqref{eq:1D_hamiltonian} are time-reversal symmetric and $\Delta \theta$ is quantized (as indicated by the horizontal dashed line).}
    \label{fig:fig5}
\end{figure}

The orbital magnetoelectric coupling describes the orbital contribution to electric polarization in response to an applied magnetic field,
\begin{align}
\alpha_{ij}
=
\frac{\partial P_i}{\partial B_j},
\end{align}
or, equivalently, the change of orbital magnetization with respect to an electric field,
\begin{align}
\alpha_{ij}
=
\frac{\partial M_j}{\partial E_i}.
\end{align}
In addition to the orbital part of the magnetoelectric response, there are spin and ionic contributions that we do not discuss further. For crystalline insulators, the orbital magnetoelectric tensor contains two distinct terms~\cite{PhysRevB.81.205104}
\begin{align}
\alpha_{ij} = (\alpha_G)_{ij} + \frac{e^2}{2\pi h} \theta_{\rm CS} \delta_{ij}.  
\label{eq:ME_alpha}
\end{align}
The first term, $\alpha_G$, is the so-called cross-gap term, which describes virtual transitions between occupied and unoccupied bands. It is an interband term; however, it is not purely geometric because it depends on details of the energy dispersion. 

The second term is isotropic and proportional to the Chern-Simons axion angle, which is given by the integral over the Chern-Simons three-form
\begin{align}\label{eq:chernsimons_3form}
    \theta_{\rm CS}
    = -\frac{1}{4\pi}\int_{T^3}
    {\rm Tr}\left[
        A dA - \frac{2i}{3}A^3
    \right]
    \quad \mod 2\pi,
\end{align}
where $A=iU^\dagger dU$ is the Berry connection for an orthonormal frame of occupied Bloch states $U=(u_1(\bm{k}),\dots,u_n(\bm{k}))$. In contrast to $\alpha_G$, the Chern-Simons term $\theta_{\rm CS}$ is purely geometric; however, it is not additive: the sum of $\theta_{\rm CS}^m$ over subsets of bands does not equal that of the full occupied bundle~\cite{PhysRevB.81.205104}. $\theta_{\rm CS}$ is $\pi$-quantized by time-reversal and/or inversion symmetry and can give rise to half-quantized surface anomalous Hall states. Time-reversal-symmetric insulators that exhibit these surfaces states are called topological insulators. The strong topological index $\nu_0$ that classifies the topology of these systems is related to the Chern-Simons terms by $(-1)^{\nu_0} = e^{i\theta_{\rm CS}}$. Insulators where the quantization of $\theta_{\rm CS}$ is protected by inversion rather than time-reversal symmetry are referred to as axion insulators~\cite{surfaces_axion_insulators}. 

Importantly, in order for the Chern-Simons three-form to be well-defined, the occupied bundle needs to admit a global frame over the BZ, i.e., the first Chern numbers of all BZ-slices are required to vanish~\cite{qi_topological_field_theory,PhysRevLett.102.146805}. This makes $\theta_{\rm CS}$ challenging to compute in the general, in particular when it is not quantized by symmetry~\cite{david_gauge_discontinuity}. Since the Chern-Simons three-form is related to the Berry curvature density by~\cite{qi_topological_field_theory}
\begin{align}
    d {\rm Tr}\left[
        A dA - \frac{2i}{3}A^3
    \right] = {\rm Tr} \left[ F\wedge F \right],
\end{align}
we can express the variation of $\theta_{\rm CS}$ w.r.t. some auxiliary parameter $\lambda_3$ through the second Chern form --- even when the first Chern numbers do not vanish. For a family of three-dimensional insulators parameterized by $\lambda_3$, the difference at two parameter values is given by  
\begin{align}\label{eq:theta_difference}
    \Delta\theta_{\rm CS}
    &=
    \theta_{\rm CS}(\lambda_3)-\theta_{\rm CS}(0)
    \nonumber\\
    &=
    -\frac{1}{16\pi}    \int_{{T^3}}     \int_{0}^{\lambda_3} d\lambda_3'\,     \mathrm{Tr}\left[        F(\bm{k},\lambda_3')\wedge         F(\bm{k},\lambda_3')    \right].
\end{align}
If the parameter $\lambda_3$ is periodic, $\lambda_3\in[0,2\pi]$, the total change of $\theta_{\rm CS}$ over one cycle is related to the second Chern number by
\begin{align}\label{eq:2ndchern_and_theta}
    \mathrm{ch}_2
    =
    -\frac{1}{2\pi}
    \int d\theta_{\rm CS}.
\end{align}

Similarly, we can recast our expression for the DDKS number in Eq.~\eqref{eq:integral_higherF_eq_nu} into
\begin{align}\label{eq:higher_2Dslices}
    \nu = \frac{1}{2\pi}\sum_{\Delta^3\in |T^3|}F(\Delta^3) = \frac{1}{2\pi} \sum_{|T^2| \in T^3}\theta_{\rm HBP}(|T^2|),
\end{align}
where we defined the higher Berry phase of two-dimensional slices through $T^3$ with constant $\lambda_3$ as
\begin{align}\label{eq:theta_higher_berry_phase}
    \theta_{\rm HBP} = \sum_{\Delta^2 \in |T^2|} \phi(\Delta^2) \mod 2 \pi,
\end{align}
with the discrete Berry connection $\phi(\Delta^2)$ from Eq.~\eqref{eq:dis_Berry_connection}. {\color{black}Here, $\theta_{\mathrm{HBP}}$ denotes the discrete iMPS higher Berry phase constructed in Ref.~\cite{shiozaki2023higher}. Its exponential is the oriented product of triangle phase factors over a closed surface. The gauge-dependent link phases cancel between adjacent triangles, yielding a gauge-invariant surface quantity that plays the role of a discrete gerbe holonomy.} Just like the boundaries of the individual tetrahedra $\partial \Delta^3$, the slices $|T^2|$ are closed two-dimensional surfaces and are thus associated with higher Berry phases that are gauge-invariant. Since we have established that $\nu = \mathrm{ch}_2$ (see Eq.~\eqref{eq:kuenneth}), the comparison of Eq.~\eqref{eq:2ndchern_and_theta} with Eq.~\eqref{eq:integral_higherF_eq_nu} is suggestive of the relationship between $\theta_{\rm CS}$ and the higher Berry phase $\theta_{\rm HBP}$. Both quantities integrate to yield the second Chern number, which raises the question if $\theta_{\rm HBP}$ and $\theta_{\rm CS}$ are identical. In following, we present numerical results on the four-dimensional Dirac model in an attempt shine light on this issue.

With Eq.~\eqref{eq:higher_2Dslices} as our inspiration, we can reinterpret the momentum $k_3$ from Eq.~\eqref{eq:tb_4d_hamiltonian} (or $\lambda_3$ from Eq.~\eqref{eq:1D_hamiltonian}) as an adiabatic pumping parameter. For any value of $k_3$, we obtain a three-dimensional lattice model or --- equivalently at constant $\lambda_3$ --- a family of one-dimensional chains parameterized over the two remaining momenta. We plot the band structures averaged over $k_4$ for $m=1$ at $k_3=\pi$ and for $m=3$ at $k_3=\pi / 5$ in Fig.~\ref{fig:fig5}(a) and (b), respectively. The spectra are symmetric w.r.t. to reflection over zero energy. Consequently, the cross-gap term $\alpha_G$ vanishes~\cite{PhysRevB.81.205104} and the orbital magnetoelectric response is determined by the Chern-Simons term alone. For this particular model, all first Chern numbers vanish such that $\theta_{\rm CS}$ can be evaluated explicitly by integrating the second Chern form in Eq.~\eqref{eq:ch2_density_dirac}. Figure~\ref{fig:fig5} shows the analytical result for the change in the Chern--Simons term, $\Delta\theta_{\rm CS}$, and the numerical result for the change in the higher Berry phase, $\Delta\theta_{\rm HBP}$. We reused the iMPS and corresponding link variables over the same triangulation presented in Sec.~\ref{sec:higher_berry}. Because inversion and time-reversal symmetry are broken within the individual slices of constant $\lambda_3$, $\Delta\theta_{\rm CS}$ is not quantized except at the time-reversal-invariant momenta (TRIM) $\lambda_3=0$ and $\lambda_3 = \pi$. We find $\Delta\theta_{\rm CS}$ to be in excellent agreement with $\Delta\theta_{\rm HBP}$. {\color{black}We note that this agreement concerns changes in the two phases
along the pumping cycle. Equality of the associated integer
invariants constrains their total winding, but does not by itself
imply agreement at intermediate parameter values. We have not
established whether this stronger agreement persists under
general gap-preserving perturbations of the Dirac model.} Next, we study a model where topology provides a guidepost for the comparison between the Chern-Simons term and the higher Berry curvature.

\section{Hopf Insulator}

{\color{black}Hopf insulators were introduced as three-dimensional two-band insulators characterized by an integer Hopf invariant~\cite{Moore2008}. Subsequent work constructed lattice models with arbitrary Hopf invariant~\cite{Deng2013} and established a correspondence between the bulk Hopf invariant and the surface Chern number~\cite{PhysRevB.103.045107}. Connections between Hopf topology, gerbe invariants, and magnetoelectric response have also been investigated~\cite{refcomment1_4}. Here, we use a Hopf-insulator model to compare the established Chern--Simons axion coupling with the higher Berry phase obtained from our chosen iMPS construction.}

\begin{figure}
    \centering
    \includegraphics[width=1\linewidth]{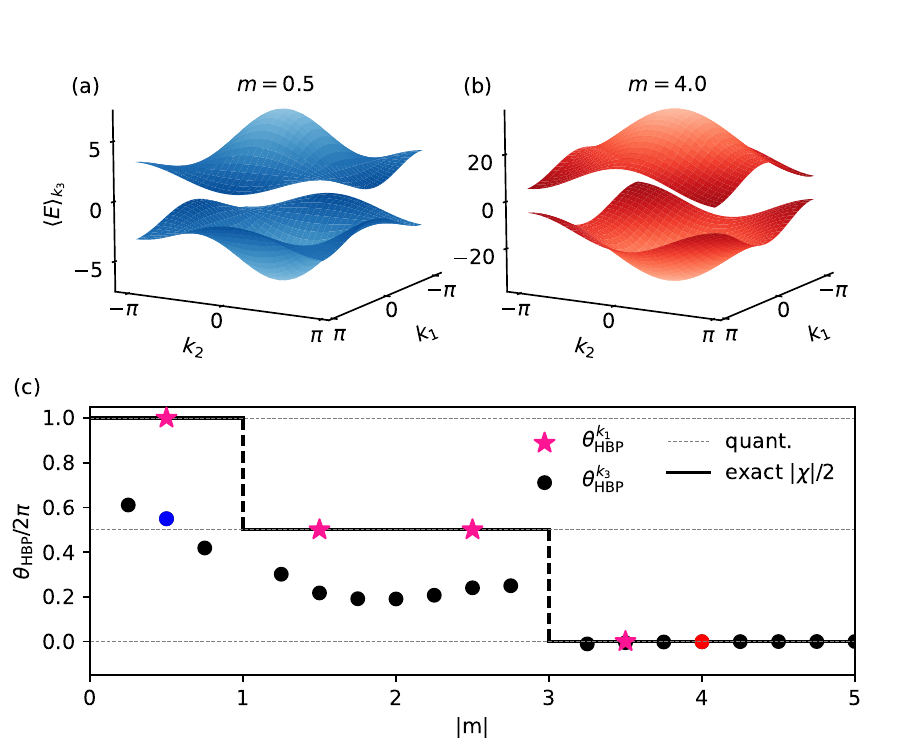}
    \caption{(a), (b) The energy density of Eq.~\eqref{eq:tb_hopf} averaged over $k_3$ with $m=0.5$ and $m=4$ as a function of $k_1$ and $k_2$ {\color{black}on a $30^2$ grid}. The spectra are symmetric w.r.t. reflection over the Fermi level. (c) $\theta_{\rm HBP}/2\pi$ for the Hopf insulator in Eq.~\eqref{eq:tb_hopf_components} as a function of $m$. The vertical solid line marks the phase transition point with a unit change in the Hopf invariant $\chi$ and a drop from finite $\theta_{\rm HBP}$ to $\theta_{\rm HBP} \approx 0$. For $|m| <3$ there is no quantization of {\color{black}$\theta^{k_3}_{\mathrm{HBP}}$}. {\color{black}$\theta_{\rm HBP}^{k_1}$ (pink stars) is quantized and agrees with the Hopf invariant.}}
    \label{fig:fig6}
\end{figure}

We consider the Hopf insulator on a lattice described by the tight-binding model~\cite{PhysRevB.103.045107}
\begin{align}\label{eq:tb_hopf}
    \mathcal{H}^{\mathrm{Hopf}}_\mathbf{k} = \psi_\mathbf{k}^\dagger \left[\mathbf{h}_\mathbf{k}\cdot \boldsymbol{\tau}\ \right]\psi_\mathbf{k},
\end{align}
with the Pauli vector $\boldsymbol{\tau} = (\tau_1, \tau_2, \tau_3)^\top$ and
\begin{align}\label{eq:tb_hopf_components}
    \mathbf{h}_\mathbf{k} = -2\begin{pmatrix} \sin k_1 \sin k_3 + \sin k_2 \cos k_3 + w_{\mathbf{k}}\sin k_2 \\  \sin k_1 \cos k_3 - \sin k_2 \sin k_3+ w_\mathbf{k}\sin k_1   \\ -w_\mathbf{k}\left(m + \cos k_3 \right) - \frac{1}{2}C_\mathbf{k} \end{pmatrix}
\end{align}
with $C_\mathbf{k} =\left(\cos k_1 + \cos k_2 \right)^2  + \left(\cos^2 k_1 + \cos^2 k_2  \right) - m^2 -1$ and $w_\mathbf{k} = \cos k_1 + \cos k_2 + m $. We plot the model's dispersion relation averaged over $k_3$ for $m=0.5$ and $m=4$ in Fig.~\ref{fig:fig6}(a) and (b), respectively. Reflecting the band structure over zero energy leaves the spectrum invariant, i.e. the cross-gap term $\alpha_G$ vanishes~\cite{PhysRevB.81.205104}. Since the model lacks both inversion and time-reversal symmetry, we naively expect $\Delta\theta_{\rm CS}$ to be non-quantized. However, we can construct a linking number (or Hopf invariant) which allows us to relate $\theta_{\rm CS}$ to the topological phase diagram of the Hopf insulator.

Because Eq.~\eqref{eq:tb_hopf} is a two-band model, its Bloch vectors $\mathbf{b}_\mathbf{k} = \mathbf{h}_\mathbf{k} / |\mathbf{h}_\mathbf{k}|$ live on the two-sphere $S^2$. The preimage of any point $\mathbf{b}_\mathbf{k}$ forms closed loops in the three-dimensional momentum space. The Hopf invariant $\chi$ measures if the loops corresponding to two distinct points on the Bloch sphere are linked or untangled. {\color{black} For a gapped two-band Hamiltonian whose occupied band has vanishing first Chern numbers on all two-dimensional BZ tori, the Hopf invariant is
\begin{equation}
    \chi=-\frac{1}{4\pi^2}\int_{T^3}d^3k\,
    \mathbf{A}(\mathbf{k})\cdot
    \bigl[\nabla_{\mathbf{k}}\times\mathbf{A}(\mathbf{k})\bigr]
    \in\mathbb{Z},
\end{equation}
where $\mathbf{A}(\mathbf{k})=
i\langle u_-(\mathbf{k})|\nabla_{\mathbf{k}}u_-(\mathbf{k})\rangle$
is the Berry connection of a smooth, periodic occupied-band eigenstate~\cite{PhysRevB.103.045107}.} Note that $\chi$ is only well-defined for two-band models (which is referred to as delicate topology). The topological phase diagram of Eq.~\eqref{eq:tb_hopf} in terms of $\chi$ is given by~\cite{Nakamura_Hopf}
\begin{align}\label{eq:chi_phase_diagram}
    \chi(m)=\left\{\begin{array}{ll}0,\quad &|m| > 3\\ 1,\quad &1 < |m| <3\\-2,\quad & |m| < 1, \\\end{array}\right. 
\end{align}
where the model becomes gapless at the phase boundaries $|m|=1$ and $|m|=3$. Formally, the Hopf invariant is calculated by integrating the Chern-Simons three-form over $T^3$, and is therefore directly related to the axion coupling through~\cite{PhysRevB.103.045107}
\begin{align}
    \theta_{\mathrm{CS}} = \pi \chi \mod 2\pi.
\end{align}
Next, we compute the ground states of the Hopf insulator using iMPS to compare $\theta_{\mathrm{HBP}}$ to Eq.~\eqref{eq:chi_phase_diagram}.

We Fourier transform the momentum component $k_3$ in Eq.~\eqref{eq:tb_hopf} to obtain a family of chains $\{\mathcal{H}^{\mathrm{Hopf}}_{\boldsymbol{\lambda}} \}$ parameterized by $\boldsymbol{\lambda}=(\lambda_1, \lambda_2)^\top \in T^2$, where --- again --- we labeled the remaining momentum components by $\lambda_n$. The index of $C_{\boldsymbol{\lambda}}$ and $w_{\boldsymbol{\lambda}}$ is relabeled as well and they now depend on the parameter tuple $\lambda_1$ and $\lambda_2$ 

The parameterized Hamiltonian is given by
\begin{widetext}
    \begin{equation}\label{eq:hopf_1D_chain}
        \mathcal{H}^{\mathrm{Hopf}}_{\boldsymbol{\lambda}} = \sum_\ell \left[ 2\left(i\sin \lambda_1 - \sin \lambda_2\right)\left( w_{\boldsymbol{\lambda}}c^\dagger_\ell d_\ell +c_\ell^\dagger d_{\ell + 1} \right)+w_{\boldsymbol{\lambda}}\left(c^\dagger_{\ell+1}c_\ell- d^\dagger_{\ell+1}d_\ell \right) + m\left( w_{\boldsymbol{\lambda}} +  \frac{C_{\boldsymbol{\lambda}}}{2m}\right)\left(c^\dagger_{\ell}c_\ell- d^\dagger_{\ell}d_\ell \right) \right] + \mathrm{h.c.}
    \end{equation}
\end{widetext}

{\color{black}Throughout, $w_{\boldsymbol{\lambda}}$ and $C_{\boldsymbol{\lambda}}$ denote the expressions defined above with their two momentum arguments replaced by $\lambda_1$ and $\lambda_2$. The identification of these parameters with the remaining momenta is specified for each choice of Fourier-transform direction.} In Fig.~\ref{fig:fig6}(c), we compare $\theta_{\mathrm{CS}}$, obtained from the topological phase diagram in Eq.~\eqref{eq:chi_phase_diagram}, with our numerical results for {\color{black}$\theta^{k_3}_{\mathrm{HBP}}$}. Unlike in the Dirac model, where the two quantities agree exactly (see Fig.~\ref{fig:fig5}) the Hopf model exhibits a clear discrepancy: for $|m|<3$, {\color{black}$\theta^{k_3}_{\mathrm{HBP}}$} is neither quantized nor equal to $\theta_{\mathrm{CS}}$. {\color{black}We note that the family of chains $\{\mathcal{H}^{\mathrm{Hopf}}_{\boldsymbol{\lambda}} \}$ is not unique, because we chose to Fourier transform a particular momentum component. Next, we consider the $k_1$- and $k_2$-families.}

{\color{black} We generate a second family of chains $\{\tilde{\mathcal{H}}^{\mathrm{Hopf}}_{\boldsymbol{\lambda}} \}$ by Fourier transforming $k_1$. We relabel the remaining momentum components $k_2 \rightarrow \lambda_2$ and $k_3 \rightarrow \lambda_1$. The $k_1$-family is given by}
\begin{widetext}{\color{black}
    \begin{align}\label{eq:hopf_k1_family}
        \tilde{\mathcal{H}}^{\mathrm{Hopf}}_{\boldsymbol{\lambda}} =- \sum_\ell  \bigg[ (&i\sin \lambda_1 + \sin \lambda_2 )\left(c^\dagger_{\ell+1}d_\ell + d^\dagger_{\ell +1}c_\ell \right) - w_{\boldsymbol{\lambda}}\left(c^\dagger_{\ell+1}c_\ell - d^\dagger_{\ell + 1}d_\ell - c^\dagger_{\ell + 1}d_\ell + d^\dagger_{\ell + 1}c_\ell \right) \nonumber \\+ 2 \sin \lambda_2(&i\sin \lambda_1 + w_{\boldsymbol{\lambda}}) c^\dagger_\ell d_\ell - \frac{1}{2}\left(c^\dagger_{\ell+2}c_\ell - d^\dagger_{\ell + 2}d_\ell - c^\dagger_{\ell + 2}d_\ell + d^\dagger_{\ell + 2}c_\ell \right)   - m\left(w_{\boldsymbol{\lambda}} + \frac{C_{\boldsymbol{\lambda}}}{2m} - \frac{\cos 2\lambda_1}{2m} \right) \left(c^\dagger_\ell c_\ell - d^\dagger_\ell d_\ell \right) \bigg] +\mathrm{h.c.}
    \end{align}}
\end{widetext}

{\color{black}Due to fourfold rotational symmetry, Fourier transforming along $k_1$ and $k_2$ yields equivalent families. Eq.~\eqref{eq:hopf_1D_chain} and Eq.~\eqref{eq:hopf_k1_family} are isospectral, i.e. the cross-gap term remains zero. We find the higher Berry phase obtained from the $k_1$-family $\theta^{k_1}_{\mathrm{HBP}}$ to be in perfect agreement with the Hopf invariant (see Fig.~\ref{fig:fig6}(c), pink stars).}

{\color{black}To understand the difference between $\theta^{k_1}_{\mathrm{HBP}}$ and $\theta^{k_3}_{\mathrm{HBP}}$, we take a closer look at the symmetries of Eq.~\eqref{eq:tb_hopf}. The Hopf insulator satisfies the particle-hole relation
\begin{align}
    \tau_x\left(\mathcal{H}^{\mathrm{Hopf}}_{k_1, k_2, k_3}\right)^* \tau_x =-\mathcal{H}^{\mathrm{Hopf}}_{-k_1, -k_2, k_3}.
\end{align}
In real space, the exchange of particles and holes between the orbitals is represented as an onsite unitary with $\mathcal{P}c_\ell\mathcal{P}^{-1} = d_\ell^\dagger$ and $\mathcal{P}d_\ell\mathcal{P}^{-1} = c_\ell^\dagger$. This unitary $\mathcal P$ maps the vacuum of the $c_\ell$ and $d_\ell$ fermions
to the fully occupied state.
For the $k_3$-family, the remaining momentum components, $k_1$ and $k_2$, both acquire a minus sign 
\begin{align}
    \mathcal{P}\mathcal{H}^{\mathrm{Hopf}}_{\boldsymbol{\lambda}} \mathcal{P}^{-1} = \mathcal{R}\mathcal{H}^{\mathrm{Hopf}}_{-\boldsymbol{\lambda}}\mathcal{R}^{-1},
\end{align}
with $\mathcal{R}$ reflecting the chain. On the other hand, only $k_2$ is reversed in the $k_1$-family, such that
\begin{align}
    \mathcal{P}\tilde{\mathcal{H}}^{\mathrm{Hopf}}_{\lambda_1, \lambda_2} \mathcal{P}^{-1} = \tilde{\mathcal{H}}^{\mathrm{Hopf}}_{\lambda_1, -\lambda_2}. 
\end{align}
The relation for the $k_3$-family preserves the orientation of the parameter-space torus, whereas that for the $k_1$-family reverses its orientation. Since the higher Berry phase is invariant under the onsite unitary $\mathcal{P}$ and changes sign under reversal of the parameter-space orientation, we obtain $\theta^{k_1}_{\mathrm{HBP}} = -\theta^{k_1}_{\mathrm{HBP}} \mod{2\pi}$ implying $\pi$-quantization of $\theta^{k_1}_{\mathrm{HBP}}$. For the $k_3$-family the symmetry involves spatial reflection of the chain, which transposes the mixed transfer matrices up to similarity transformations and maps the higher Berry connection constructed from right eigenvectors to one constructed from left eigenvectors; as discussed in Ref.~\cite{shiozaki2026equivariant}, no general transformation law relating these connections is established, so the preceding onsite-symmetry argument does not enforce quantization of $\theta^{k_1}_{\mathrm{HBP}}$. The agreement between $\theta_{\mathrm{CS}}$ and $\theta_{\mathrm{HBP}}$ in both the Dirac and the Hopf model provides encouraging evidence for a connection between the two angles. We discuss the implications of these results in the conclusion.

}

\section{\label{sec:conclusion}Conclusion}

We have mapped the four-dimensional Chern insulator to a family of infinite one-dimensional Hamiltonians and have computed the higher three-form Berry curvature across the three-dimensional parameter space using their iMPS ground states. We have demonstrated that the corresponding DDKS number $\nu$ generates the same topological phase diagram as the second Chern number $\mathrm{ch}_2$, which we have computed numerically for comparison. We conclude this manuscript by discussing potential experimental implications of nontrivial DDKS numbers and try to position our results in a broader context of quantum geometric and topological properties.

In the language of Thouless pumps~\cite{thoulesspump}, a finite first Chern number refers to the number of elementary charges pumped within one period, whereas the DDKS number measures ``how many Chern numbers'' are pumped across one pumping cycle. In two dimensions, it is well-known that the first Chern number counts the number of chiral edge states on the boundary of a two-dimensional sample as mandated by the Thouless-Kohmoto-Nightingale-den Nijs bulk-boundary correspondence\cite{TKNN}. In four dimensions, the second Chern number governs the magnetoelectric response, which we have shown to be related to the DDKS number using the K\"unneth formula. Since condensed matter systems are typically three- or lower-dimensional, systems with synthetic dimensions seem to be promising experimental platforms to study higher Berry physics~\cite{syn_review}. In these setups, an additional internal degree of freedom --- which usually comprises atomic states or excitation levels of resonators --- is exploited to serve as an additional lattice dimension\cite{syn_science1,syn_science2}. Four-dimensional Hall effects have been experimentally realized in cold atom and integrated photonics systems\cite{4d_hall_cold_atoms,synthetic_carusotto_photonics}. Another recent study proposes that higher-dimensional pumps can be realized by quasi-periodic driving of wires with different frequencies\cite{quasi-periodic_drive}, and collections of chains of qubits with diabatically changing couplings can, in principle, be realized on existing quantum annealing platforms\cite{king2024computationalsupremacyquantumsimulation}. In systems with dimension lower than four, it was proposed that the second Chern number can be probed through microwave spectroscopy in mesoscopic superconductors~\cite{ch2_sc}, or that pair-density states in three-dimensional topological semimetals might feature nonabelian Majorana modes protected by $\mathrm{ch}_2$~\cite{PhysRevLett.118.207002}.  {\color{black}A connection between higher Berry curvature and nonlinear shift-current responses in three dimensions has been
established using tensor Berry connections~\cite{refcomment1_4}.
Whether an analogous relation holds for the DDKS higher Berry
curvature considered here remains to be established.}

Following the dimensional hierarchy of topological invariants, it is natural to ask if there is a relation between the higher Berry phase and magnetoelectricity in three-dimensions --- in particular through the Chern-Simons axion coupling $\theta_{\mathrm{CS}}$. For the four-dimensional Chern insulator, the non-quantized change of $\theta_{\mathrm{CS}}$ along the pumping cycle is reproduced precisely by the higher-Berry-phase angle. This provides a direct realization of the expected connection between the second Chern number in the extended parameter space and the winding of the Chern--Simons magnetoelectric response. {\color{black}For the Hopf model, an appropriate choice of Fourier-transform direction likewise yields $\theta_{\mathrm{HBP}}=\theta_{\mathrm{CS}} \mod{2\pi}$. However, not every chain family obtained from the same lattice Hamiltonian produces a quantized higher Berry phase, highlighting the importance of the symmetry properties of the resulting family. While our findings do not prove a general equivalence or establish that an appropriate iMPS family always exists, the agreement found in both models is highly encouraging and motivates further research into the conditions under which higher Berry phases capture the Chern–Simons coupling.}

{\color{black}Beyond the proposed correspondence with the Chern–Simons angle, higher Berry phases offer a geometric characterization of quantum states that may admit broader physical interpretations. Connections between higher Berry curvature and magnetoelectric and nonlinear optical responses have been established within a different tensor Berry connection framework~\cite{refcomment1_4}. Our results therefore motivate clarifying how the particular
iMPS construction employed here relates to these response-based
formulations.} Given the recent interest in geometric contributions to nonlinear transport~\cite{fu,Ma2019,jenn,Lai2021,science_nonlinear,yannis}, ground-state physics~\cite{LLspread,vortexability1,ledwith2024nonlocalmomentschernbands}, and quantum fluctuations~\cite{taisei,onishi2024quantumweight,qm_altermagnet}, we expect that the higher Berry phase may eventually provide a useful diagnostic information, even {\color{black} beyond its relation to} $\theta_{\mathrm{CS}}$. Related higher-form structures in single-particle band theory have been discussed in Refs.~\cite{tomas,PhysRevB.99.045154}, although the DDKS invariant considered here arises from a different many-body construction.

While we focused on the classification of free fermions in this manuscript, iMPS are naturally designed to be an efficient representation of even strongly interacting quantum states. We are hopeful that our method will prove useful to study the topology of interacting fermionic systems in the future. To lowest order, a mean-field treatment --- e.g. of a local density-density repulsion --- leads to a mass-term renormalization through which a topological phase transition can be driven (see Eq.~ \eqref{eq:ch2_phase_diagram} or Eq.~ \eqref{eq:chi_phase_diagram}). If such interactions have higher order effects like a redistribution of higher Berry curvature across parameter space is not known, but plausible. We want to stress that adding a local interaction (like a ``Hubbard-$U$") to the four-dimensional Dirac Hamiltonian in Eq.~\eqref{eq:1D_hamiltonian} does \textit{not} describe the physics of the interacting four-dimensional Chern insulator, because the correlations are constrained to spread only along the real-space dimension, but not across parameter space. A suitable material platform would consist of a collection of (quasi) one-dimensional systems with screened interactions between the chains. The organic superconductors\cite{organic_sc} or  Li$_{0.9}$Mo$_6$O$_1$\cite{limoo} are a class of such quasi one-dimensional systems. Other potential candidates include the interacting Su-Schrieffer-Heeger model~\cite{yoshida,kyle}, spin chains with additional unit-cell degrees of freedom~\cite{tensnet_edge1,efficient_moire,tensnet_edge2,tensnet_edge3} or generalizations of the Kitaev $p$-wave chain~\cite{kitaev2001unpaired} -- so-called parity pumps~\cite{parity_pump}. 

Regardless of the presence or absence of many-body interactions, only very recently it has become possible to numerically compute higher Berry curvature for generic models at all\cite{shiozaki2023higher,sommer1,sommer2}. Consequently, the number of platforms with known nontrivial DDKS numbers is still small. We consider it vital to identify more examples and hope that our work inspires more research on that frontier. 

\textit{Note added.---} Upon finalizing this paper, we became aware of related works~\cite{cole2026exactexpressionberryconnection, lo2026detectinghigherberryphase}. The former proposes an efficient numerical method for computing the magnetoelectric axion coupling without the use of finite differentiation. The latter discusses the fact that the higher invariant for free fermions can be extracted from the parameter-dependent winding number of the boundary scattering matrix.
In addition, Ref.~\cite{singam2026texturedphasediagramsfeatureless} discussed the relationship between the Dixmier--Douady class and the second Chern number in free-fermion models. Some of the results presented in this work were previously reported in the first author's doctoral thesis~\cite{Heinsdorf_2025}. 

\begin{acknowledgments}
We thank Aris Alexandradinata for helpful discussions {\color{black} and Joel Moore for stimulating input that led us to find agreement between the higher Berry phase and the Chern–Simons axion angle in the Hopf insulator}. NH acknowledges support through the ``Quantum Electronic Science and Technology" program awarded by the Quantum Matter Institute at the university of British Columbia, Canada, as well as computational resources and services provided by Advanced Research Computing at the University of British Columbia. NH also acknowledges financial support through the Japanese Society for the Promotion of Science (JSPS) and the German Academic Exchange Service (DAAD) within the JSPS Summer Program 2020 (postponed to 2022 due to COVID-19 pandemic) and H. Katsura through which the initial contact between the authors was established. NH further acknowledges financial support by the Max Planck Institute for Solid State Research in Stuttgart, Germany for traveling that enabled discussion between the authors. NH also acknowledges financial support through the Feodor-Lynen fellowship awarded by the Humboldt foundation. KS were supported by JST CREST Grant No. JPMJCR19T2, and JSPS KAKENHI Grant No. JP22H05118 and JP23H01097.
\end{acknowledgments}

\appendix

\section{Comparison to Analytic Berry Curvature Density\label{app:analytic_FF}}
The analytic expression for the second Chern number of Eq.~\eqref{eq:tb_4d_hamiltonian} is known\cite{qi_topological_field_theory}. We first define
\begin{align}
    \mathbf{d}(\mathbf{k}) = \left( m + \sum_\mu \cos k_\mu, \sin k_x, \sin k_y, \sin k_z, \sin k_w \right),
\end{align}
and the Bloch vector 
\begin{align}
    \mathbf{b}(\mathbf{k}) = \mathbf{d}(\mathbf{k}) / E_+(\mathbf{k}).
\end{align}
The second Chern number is then given by
\begin{align}\label{eq:ch2_density_dirac}
    \mathrm{ch}_2 = \frac{3}{8\pi^2}\int d^4k \epsilon^{abcde} b_a(\mathbf{k})\partial_{k_1} b_b(\mathbf{k})\partial_{k_2} b_c(\mathbf{k})\partial_{k_3} b_d(\mathbf{k})\partial_{k_4} b_e(\mathbf{k}).
\end{align}
Naturally, this expression can be evaluated on a much finer grid than the higher Berry curvature using iMPS. We compute the integrand on a $50^4$-grid, integrate out $k_4$, and then apply a $5^3$ summation mask with stride size $5$ to obtain the Berry curvature density on a $10^3$ grid that we can compare to the higher Berry curvature $F$. 

\bibliographystyle{apsrev4-2}
\bibliography{bibliography}

\end{document}